 \documentclass[aps,prx,twocolumn,superscriptaddress,longbibliography]{revtex4-1}
\usepackage{titlesec}
\usepackage{amsmath}  
\usepackage{amsfonts} 
\usepackage{graphicx}
\usepackage{amssymb}
\usepackage{hyperref} 
\hypersetup{colorlinks=true,urlcolor=blue,citecolor=blue,linkcolor=blue}
\usepackage{cleveref}
\usepackage{etoolbox}
\AtBeginEnvironment{quote}{\singlespacing\small}

\DeclareMathOperator{\dd}{d\!}
\DeclareMathOperator{\ddd}{\mathrm{d}}

\begin{document}

\title{Demystifying the Lagrangian of classical mechanics} 

\author{Gerd Wagner}
\affiliation{Mayener Str. 131, 56070 Koblenz, Germany} 
\email{\\gerdhwagner@t-online.de} 


\author{Matthew W. Guthrie}
\affiliation{Department of Physics, University of Connecticut, Storrs, CT USA}
\email{\\guthrie@uconn.edu}


\date{\today}


\begin{abstract}
The Lagrangian formulation of classical mechanics is widely applicable in solving a vast array of physics problems encountered in the undergraduate and graduate physics curriculum. 
Unfortunately, many treatments of the topic lack explanations of the most basic details that make Lagrangian mechanics so practical. 
In this paper, we detail the steps taken to arrive at the principle of stationary action, the Euler-Lagrange equations, and the Lagrangian of classical mechanics.
These steps are: 
1) we derive the Lagrange formalism purely mathematically from the problem of the minimal distance between two points in a plane, introducing the variational principle and deriving the Euler-Lagrange equation;
2) we transform Newton's second law into an Euler-Lagrange equation, proving that the Lagrangian is kinetic minus potential energy;
3) we explain why it is important to reformulate Newton's law.
To do so, we prove that the Euler-Lagrange equation is astonishingly the same in any set of coordinates.
We demonstrate that because of this feature the role that coordinates play in classical mechanics is much simpler and clearer in the Lagrangian as compared to the Newtonian formulation. 
This is important because the choice of coordinates is not relevant to physical reality, rather they are arbitrarily chosen to provide a convenient way of analyzing a physical system.


\end{abstract}

\maketitle

\section{Introduction}\label{introduction}

\subsection{The classical Lagrangian in recent literature}

The Lagrangian formalism is a powerful description of classical mechanics~\cite{lagrange1811}. 
To some, however, Lagrangian mechanics can be seemingly separate from the more familiar Newtonian formulation, and many struggle with understanding how and why the Lagrangian formulation is an equivalent description of motion~\cite{Hopkins2011, Bouzounieraki}. 
This is understandable, as Lagrangian mechanics presents an entirely new way to think about physics and this shift in thinking can be intellectually taxing to overcome, especially in the context of a university course.
Often through a lack of class time to cover material, many treatments of Lagrangian mechanics in university courses and popular textbooks rely on incomplete arguments, especially when considering the Lagrangian ($L$) itself. 
It is especially difficult to understand why the Lagrangian takes the intriguing form $L=T-V$, the difference between a particle's kinetic ($T$) and potential ($V$) energies (which are functions of position $r$, velocity $\dot{r}$, and time $t$). 
We aim to address all of these issues in the current paper. 

Our methods differ significantly from the most common undergraduate level treatments of the Lagrangian formulation, many of which simply define the Lagrangian functional and move on to using it to solve problems.
Marion~\cite[p.~198]{marion1970classical} as well as Fowles and Cassiday~\cite[p.~393]{fowles1999analytical} merely define the Lagrangian and move to its utility. 
Taylor~\cite[p.~238]{taylor2005classical} proceeds similarly, but includes the explicit statement that the reader is ``certainly entitled to ask why the quantity $T-U$ should be of any interest''\footnote{Taylor uses $U$ for the potential energy function.} and continues on to say that ``there seems to be no simple answer to this question except that it is.''
Morin~\cite[p.~218]{morin2008introduction} directly addresses the curious sign of the potential energy: ``Yes, there is a minus sign in the definition (a plus sign would simply give the total energy)'' but does not explain \textit{why} that minus sign appears!
Gregory~\cite[p.~348]{gregory2006classical} explains the Lagrangian in further detail, including a detailed derivation, although the information is presented over multiple sections and chapters with numerous digressions.
Similarly, while neither recent nor written for the undergraduate level, it is important to note that Landau \& Lifshitz~\cite[p.~2-10]{landauLifshitz1976Mechanics} presume ``the principle of least action'' as an axiom and then utilize several facets of nature to derive the Lagrangian. 

\subsection{Motivation for an alternative point of view}\label{sec:motivation}

The Lagrangian formalism arises naturally through an introductory example from geometry in our treatment presented in this paper. 
We then utilize the formalism to reformulate Newton's second law in the form of the Euler-Lagrange equation. 
To motivate the new formulation of Newton's second law, we prove the invariance of the Euler-Lagrange equation under arbitrary coordinate transformations. 
To make clear why this property is important for physics, we remind the reader that the laws of nature do not depend on the coordinates we use to describe them. 
On the other hand, physicists cannot formulate laws without coordinates, and physical equations usually look different in different coordinates~\footnote{Coordinate-free laws of physics do exist (see Maxwell's equations written in the language of differential forms or Einstein's equations). However, these formulations are not very useful until a coordinate system is chosen in which to apply them.}. 
This is why the relationship between physical theories and coordinates should be as well-defined and restricted as possible. 
The Lagrangian formalism fulfills this demand through the Euler-Lagrange equations being independent of coordinate transformations. 
Lagrangian functionals do depend on coordinates, but in the simplest way physicists can think of: they transform like scalars.

Thinking about the relation of the laws of physics to coordinates proved fruitful in the past.
Probably the most famous example is Einstein's first paper on special relativity~\cite{EinsteinSpecialRelativity, einstein1905electrodynamics}.
The current paper follows this tradition, reformulating Newton's second law in the form of Euler-Lagrange equations. 
As a result, the Lagrangian of classical mechanics appears naturally without requiring further motivation or derivation.

\section{The Lagrangian formalism and the minimal distance between two points in a plane}\label{distance}

Because it is a simple introduction to both the functional that we will derive to be the Lagrangian, and to techniques in the calculus of variations (a pre-requisite topic to understanding the Lagrangian formalism), we aim to show that the minimal distance between two points in a plane is a straight line~\footnote{Some limitations are: we limit ourselves to lines that can be written as one dimensional functions $y=f(x)$ while curves in two dimensions would be more general. 
We assume the shortest distance must be a differentiable function (although continuous would be sufficient). 
Foremost, we satisfy ourselves with finding a condition that makes the curve only stationary instead of minimal.}. 

The arc length $S$ of a function $y(x)$ between two points $(x_1,y_1)$ and $(x_2,y_2)$ is given by
\begin{equation}
S=\int_{x_1}^{x_2}\sqrt{\dd x^2 + \dd y^2}.
\end{equation}
Factoring $\dd x$ from the radical results in the following equation
\begin{equation}
S= \int_{x_1}^{x_2}\sqrt{1 + \left(\frac{\dd y}{\dd x}\right)^2} \dd x,
\end{equation}
and defining $y' := \frac{\dd y}{\dd x}$ gives a convenient representation for the equation representing the arc length of $y(x)$ as a function of how $y$ changes over its length, 

\begin{equation}
S = \int_{x_1}^{x_2}\sqrt{1 + y'^2} \dd x.
\end{equation}
We generalize this formula by writing
\begin{equation}
S=\int_{x_1}^{x_2} G(y,y',x) \dd x .
\end{equation}
Although in our example $G = \sqrt{1 + y'^2}$ only depends on $y'$ and not explicitly on $y$ or $x$, we can also assume dependence on $y$ and $x$. 
If formulas we derive for $G$ contain derivatives of $G$ with respect to $y$ or $x$, we just replace these terms with zero as the derivative of a function with respect to a variable it does not depend on is always zero.


To find the function $y(x)$ that minimizes $S$, we make the simplifying assumption that it is sufficient to find the $y(x)$ that makes $S$ stationary. 
In other words, we assume that $S$ has exactly one minimum.
To do this, we consider small but arbitrary variations $\delta y$ of $y$ and try to find a condition that causes $\delta S$ to vanish. 
These variations allow the path taken by the function to change by a small amount.
During this, the endpoints $(x_1,y_1)$ and $(x_2,y_2)$ are kept fixed; therefore the variations $\delta y$ have the property $\delta y(x_1) = \delta y(x_2) = 0$. 
As a result,

\begin{equation}
\delta S = \int_{x_1}^{x_2} \left(\frac{\partial G}{\partial y} \delta y
+ \frac{\partial G}{\partial y'} \delta y' \right) \dd x,
\end{equation}
where $\frac{\partial G}{\partial y}$ indicates the partial derivative of $G$ with respect to $y$, the partial derivative being necessary because $G$ is a function of both $x$ and $y$. 
Considering $\delta y$ as the difference of two functions $\delta y(x) = y_b(x) - y_a(x)$ we can write
$\delta y' = y_b' - y_a' = \frac{\ddd }{\dd x}(y_b - y_a) = \frac{\ddd}{\dd x} \delta y$. With this and by using integration by parts for the second term, we find
\begin{equation}
\delta S = \int_{x_1}^{x_2} \left( \frac{\partial G}{\partial y} \delta y
- \frac{\ddd}{\dd x}\frac{\partial G}{\partial y'} \delta y \right) \dd x
+ \left[\frac{\partial G}{\partial y'} \delta y \right]_{x_1}^{x_2}.
\end{equation}
The last term vanishes because $\delta y(x_1) = \delta y(x_2) = 0$, and we are left with
\begin{equation}
\delta S = \int_{x_1}^{x_2} \left( \frac{\partial G}{\partial y}
- \frac{\ddd}{\dd x}\frac{\partial G}{\partial y'} \right) \delta y \; \dd x.
\end{equation}
For $S$ to be stationary, $\delta S$ must vanish. 
Since $\delta y$ is arbitrary, the condition must be
\begin{equation}\label{e-l}
\frac{\partial G}{\partial y} - \frac{\ddd}{\dd x}\frac{\partial G}{\partial y'} = 0.
\end{equation}
Equation \eqref{e-l} is called an Euler-Lagrange equation. 
In classical mechanics, the quantity $S$ has units of energy times time and is called action, and the procedure of looking for a condition that makes $S$ stationary under a function $G(y,y',x)$ is called the Lagrangian formalism. 
If we substitute $G = \sqrt{1+y'^2}$ into Eq. \eqref{e-l},
\begin{equation}
0 - \frac{\ddd}{\dd x}\frac{y'}{\sqrt{1+y'^2}} = 0
\end{equation}
we see that the solution of the differential equation is
\begin{equation}
y(x) = c_1 x +c_2,
\end{equation}
with $c_1$ and $c_2$ as arbitrary constants. Thus, $y(x)$ is a straight line connecting $(x_1,y_1)$ and $(x_2,y_2)$.

For a more complete introduction to the calculus of variations in a physics context, see Goldstein's Classical Mechanics textbook~\cite{goldstein2002classical}, or Boas's Mathematical Methods book~\cite{boas2006mathematical}.

\section{Application to Newtonian mechanics} \label{application}
As we discussed in section \ref{sec:motivation}, it is desirable to write Newton's law $F=ma$ in the form of an Euler-Lagrange equation. 
Doing so will provide us with a function like $G$ from the previous section, only now the function will have important physical implications. 
This function is called the Lagrangian of classical mechanics.

To proceed, we rearrange Newton's second law
\begin{equation}
0 = ma - F.
\end{equation}
The first term can be rewritten as follows:
\begin{equation}
ma = m \ddot{r} = \frac{\ddd}{\dd t} (m \dot{r})
= \frac{\ddd}{\dd t} \frac{\partial}{\partial \boldmath\dot{r}\unboldmath} \left(\frac{1}{2} m \dot{r}^2 \right)
= \frac{\ddd}{\dd t} \frac{\partial T}{\partial \dot{r}},
\end{equation}
where $T:=\frac{1}{2} m \dot{r}^2$ is the classical kinetic energy of the system. 
For the second term, we assume the force $F$ to be conservative. 
Consequently, there exists a potential $V$ such that
\begin{equation}
F = - \frac{\partial V}{\partial r} = \frac{\partial (-V)}{\partial r}.
\end{equation}
Using both rewritten terms, Newton's law becomes
\begin{equation}
0 = \frac{\ddd}{\dd t} \frac{\partial T}{\partial \dot{r}} - \frac{\partial (-V)}{\partial r}.
\end{equation}
If we assume that $\partial T/ \partial r = 0$ and $\partial V / \partial \dot{r} = 0$, which is nearly always true in Newtonian mechanics, we can convert the equation to
\begin{equation}\label{e-lwithtv}
0 = \frac{\ddd}{\dd t} \frac{\partial (T-V)}{\partial \dot{r}} - \frac{\partial (T-V)}{\partial r}.
\end{equation}

\section{Invariance of the Euler-Lagrange equation under coordinate transformations} \label{invariance}

As announced in section \ref{sec:motivation} we now prove the astonishing invariance of the Euler-Lagrange equations under arbitrary coordinate transformations.

Let $y=f(Y,x)$ be an invertible and differentiable coordinate transformation.
\footnote{The most effective tool in a physicist's toolbox when solving physics problems is picking an appropriate coordinate system.
A pendulum in Euclidean $(x,y)$ coordinates makes analyzing the problem cumbersome, but in circular $(r,\theta)$ coordinates, analysis becomes simple. 
Translating from the coordinate system of the problem statement to the coordinate system that best simplifies the system usually provides great insight, but translating back to the coordinate system of the problem statement is still necessary to solve the problem.}
%
We define the transformed function $\widetilde{G}$ through $G$ by
\begin{equation} \label{lagrangian-transform}
\widetilde{G}(Y,Y',x) := G(f,f',x).
\end{equation}
Using the derivation from section \ref{distance}, we find that making $S = \int_{x_1}^{x_2} \widetilde{G}(Y,Y',x) \dd x$ stationary requires the Euler-Lagrange equation

\begin{equation}
\frac{\partial \widetilde{G}}{\partial Y}
- \frac{\ddd}{\dd x}\frac{\partial \widetilde{G}}{\partial Y'} = 0
\end{equation}
be satisfied. 
This result is again reached by considering a small but arbitrary variation $\delta Y$ which again vanishes at its endpoints.

Likewise, the variation of the same $S$ can be expressed by
\begin{equation}
\delta S = \int_{x_1}^{x_2} \left( \frac{\partial G}{\partial f} \delta f
+ \frac{\partial G}{\partial f'} \delta f' \right) \dd x
\end{equation}
where $\delta f$ is given by $\delta f = \frac{\partial f}{\partial Y} \delta Y$.

Using $\delta f' = f_2' - f_1' = \frac{\ddd}{\dd x}(f_2 - f_1) = \frac{\dd}{\dd x} \delta f$ and integration by parts for the second term we find
\begin{equation}
\delta S = \int_{x_1}^{x_2} \left( \frac{\partial G}{\partial f}
- \frac{\ddd}{\dd x} \frac{\partial G}{\partial f'} \right) \delta f \, \dd x \;
+ \; \left[\frac{\partial G}{\partial f'} \delta f \right]_{x_1}^{x_2}.
\end{equation}
As $\delta Y$ goes to $0$ at the endpoints, so does $\delta f$ which causes the last term to vanish. 
From the arbitrariness of $\delta Y$ follows the arbitrariness of $\delta f$. 
The only mechanism for $\delta S$ to vanish is
\begin{equation}
0 = \frac{\partial G}{\partial f} - \frac{\ddd}{\dd x} \frac{\partial G}{\partial f'}
= \frac{\partial G}{\partial y} - \frac{\ddd}{\dd x} \frac{\partial G}{\partial y'}.
\end{equation}
This shows that the Euler-Lagrange equation takes the same form under any coordinate transformation as long as the transformation of the function $G$ is given by $\widetilde{G}(Y,Y',x) := G(f,f',x)$. 
Functions that behave this way under coordinate transformations as $G$ are called scalar functions, especially in physical contexts.~\footnote{An example for a scalar is air temperature.
To analyze this example we consider two coordinate systems with coordinates $x$ and $X$ and their transformation $x=f(X)$.
If $T=T(x)$ denotes temperature in the first coordinate system then as an analogue to definition (\ref{lagrangian-transform}) we define the temperature in the second coordinate system by
\begin{equation} \label{definitionOfTemperatureTransform}
  \tilde{T}(X) := T(f(x)).
\end{equation}
Now the physical fact that temperature is the same in both coordinate systems leads to the condition $\tilde{T}(X) = T(x)$ which by using definition (\ref{definitionOfTemperatureTransform}) can be turned into $T(f(x)) = T(x)$.
Scalars that fulfill this condition are called invariant with respect to the transformation $f$.
An interesting point of the temperature example is that this condition becomes invalid if one of the two coordinate systems moves with a velocity that is not negligible compared to the average motion of the air molecules while the other stays relative to the air at rest.
}
In section \ref{lagrangian-def}, $G$ will be interpreted as the Lagrangian we mentioned in section \ref{introduction}. 
As we now see, $G$ does indeed have the transformation properties we claimed in section \ref{introduction}.
For a textbook example of the material presented in this section, see Ref.~\cite{hagen2009path}.

\section{Definition of the Lagrangian}\label{lagrangian-def} 

Looking back at the original Euler-Lagrange equation \eqref{e-l} and its transformed version in Eq. \eqref{e-lwithtv}, let time $t$ take the place of a general independent variable $x$, position $r$ take the place of a general coordinate $y$, and $T-V$ take the place of $G$, we arrive at the following results:

Newton's second law takes the form
\begin{equation}
0 = \frac{\ddd}{\dd t} \frac{\partial L}{\partial \dot{r}} - \frac{\partial L}{\partial r},
\end{equation}
where
\begin{equation}
 L := T-V.
\end{equation}
This $L$ is precisely the Lagrangian from classical mechanics. It is not some divinely sanctioned quantity; it is just Newtonian mechanics under a simple change of variable.

Using this representation of $L$ allows the Lagrangian formulation of mechanics to be derived by requiring that the trajectory $r(t)$ which a particle takes between two endpoints $(t_1,r_1)$ and $(t_2,r_2)$ makes the integral
\begin{equation}\label{eqref:action}
S=\int_{t_1}^{t_2} L \; \dd t
\end{equation}
stationary. 
From this, the equations of motion for the particle follow, being the particle's Euler-Lagrange equation. 
This is called the principle of stationary action in physics~\footnote{The principle of stationary action is also sometimes called the principle of \emph{least} action. This can be confusing because the Euler-Lagrange equation finds instances where action is stationary (e.g. a saddle point)~\cite{gray2007action}.}.
It is unnecessary to speciously explain the form of the classical Lagrangian as having an innate property of interest. We have shown that the Lagrangian takes its classical form
\begin{equation}
  L=T-V
\end{equation}
out of convenience.

Since $F=ma$ is formulated in Cartesian coordinates, the Lagrangian $L$ we derived is formulated in Cartesian coordinates, as well. 
The transformation law \eqref{lagrangian-transform} we found for $G$ provides us with a well-defined method for the transformation of $L$ to other coordinate systems. 
The equations of motion, which are now Euler-Lagrange equations, are likewise the same in any coordinate system.  
This way, the relationship between Newton's law and coordinates is restricted and well-defined in the sense we mentioned in section \ref{sec:motivation}. 

\section{Conclusion and Outlook}

Although the principle of stationary action is a novel interpretation of classical mechanics, it nonetheless leads to the same equations of motion as Newtonian physics.
In sections \ref{distance} and \ref{invariance}, we explored the mathematical structure of the Lagrangian formalism and, because of its transformation properties, found it desirable for classical mechanics. 
This motivated and enabled us to give comprehensive derivations of the Lagrangian of classical mechanics $L=T-V$ and the principle of stationary action. 
Our hope is that we have clarified at least one aspect of the Lagrangian formulation of classical mechanics for students who are curious about the subject, and for instructors who wish for a new way to introduce several important concepts into their curriculum.

While the transformation properties of the Lagrangian and the Euler-Lagrange equations are already reason enough to formulate physical laws using the principle of stationary action, these are by far not the sole reasons. 
The principle of stationary action has enormous analytical capabilities which lie far beyond those of Newtonian mechanics. 
As an outlook we mention some of these analytical capabilities.

\subsection{Conservation laws}
The principle of stationary action allows derivation of conservation laws from transformation properties of the Lagrangian~\cite{kleinert2016particles}. 
From this, energy, momentum, angular momentum, and other quantities that may be conserved are given formulations which depend only on the Lagrangian, coordinates, and time. 
Thus, concepts like energy, momentum, and angular momentum gain well-defined meaning for any physical system that has a Lagrangian. 
Since every physical theory can be described by the principle of stationary action (for examples see section \ref{field.theory}), the concepts of energy, momentum, angular momentum, and other conserved quantities are consistently defined for all of these theories.

An advanced treatise of energy and momentum conservation  based on the Lagrangian formalism can be found in Ref.~\cite{Ootsuka_2015}.
This paper also contains a geometrical interpretation of the Lagrangian formalism, including examples for the application of the formalism to several fields of theoretical physics, such as relativistic particle physics, relativistic field theories, and general relativity.

\subsection{Quantum mechanics}
The rules of quantum mechanics are mostly based on Hamiltonian physics and Poisson brackets, both of which are continuations of the principle of stationary action~\cite{Schwabl}. 
A recent and in-depth discussion of a Lagrangian underlying the Schr{\"o}dinger equation can be found in Ref.~\cite{Deriglazov}.

\subsection{Field theories} \label{field.theory}
The principle of stationary action has been extensively continued to field theories~\cite{Sterman,Dirac}. 
For example, Maxwell's equations can be derived from a principle of stationary action in electrodynamics. 
There, Maxwell's equations play a similar role as Newton's laws did in this paper.

Most physical theories are field theories, and as a result the widest variety of Lagrangians are field Lagrangians. 
The most common field Lagrangians, apart from that of electrodynamics, are
\begin{itemize}
  \item the Dirac Lagrangian for the relativistic field of fermions,
  \item the Klein-Gordon Lagrangian for the relativistic field of bosons,
  \item the Schr{\"o}dinger Lagrangian for non-relativistic quantum mechanics,
  \item the Lagrangian of the Standard Model of particle physics, and
  \item the general relativity Lagrangian for Einstein's theory of general relativity.
\end{itemize}

\subsection{Constrained motion}
The simplest example of a problem of constrained motion is that of particle on an inclined plane in a uniform gravitational field~\cite{kuypers2016klassische}. 
There are, of course, situations with more complex constraints than an inclined plane.
In these cases it can be difficult to find the equations of motion for the particle using Newton's laws. 
At this point, the transformation law we derived in section \ref{invariance} becomes very helpful: assume there is a transformation of coordinates that well suits the constraints to the coordinates in which the Lagrangian is formulated. 
Once that is done, Eq.~\eqref{lagrangian-transform} can be used to rewrite the Lagrangian in the coordinates that suit the constraints. 
The equations of the constrained motion are then the Euler-Lagrange equations of the transformed Lagrangian.

\subsection{Thermodynamics and statistical mechanics}
A Lagrangian approach to classical thermodynamics where the first and second law of thermodynamics are derived using the Lagrangian formalism is presented in Ref.~\cite{Stokes_2017}.
While Stokes derived those results directly from classical mechanics, there also exist more sophisticated approaches, one of which is shown in Ref.~\cite{Zee2010}.
There, the partition function of statistical mechanics~\footnote{The partition function governs statistical mechanics, an explanation of which can be found in texts on thermal physics, one common example is in Schroeder's Introduction to Thermal Physics~\cite{schroeder2021introduction}.} is shown to be related to the path integral formulation of quantum field theory.
Furthermore, the path integral formulation of quantum field theory is based on the Lagrangian formalism (also explained in Ref.~\cite{Zee2010}).

\acknowledgments{We would like to thank the many friends and mentors who helped us clarify numerous points in this manuscript and encouraged the curiosity that sparked this writing. Namely, John Jaszczak, Zhongzhou Chen, Tony Szedlak, Dan Miller, David McGhan, and Chris Riley.}

\bibliographystyle{iopart-num-mod}
\bibliography{lagrange}

\end{document}